\documentclass[conference]{IEEEtran}
\IEEEoverridecommandlockouts
\usepackage[utf8]{inputenc}
\usepackage{graphicx} 
\usepackage{bm}
\usepackage{amsmath}
\usepackage{amsfonts}
\usepackage{algorithm}
\usepackage{algpseudocode}
\usepackage{multirow}
\usepackage{amssymb}
\usepackage{pifont}
\usepackage{subfigure}
\usepackage{soul}
\usepackage{tabularx}
\newcolumntype{Y}{>{\centering\arraybackslash}X}
\newcolumntype{K}[1]{>{\centering\arraybackslash}m{#1}}
\newcolumntype{L}[1]{>{\centering\arraybackslash}p{#1}}
\usepackage{multirow}
\usepackage{xcolor}
\usepackage{marginnote}
\usepackage{subfigure}
\usepackage{cite}
\usepackage{titlesec}

 \titlespacing*{\subsection}
{0pt}{ .1ex plus .1ex minus .2ex}{ .3ex plus .2ex}

\RequirePackage{enumitem}

\setlist[enumerate]{itemsep=0pt, topsep=0pt, itemindent=15pt, leftmargin=0pt,listparindent=\parindent}
\setlist[itemize]{itemsep=0pt, topsep=0pt, itemindent=15pt, leftmargin=0pt,listparindent=\parindent}

\def\BibTeX{{\rm B\kern-.05em{\sc i\kern-.025em b}\kern-.08em
    T\kern-.1667em\lower.7ex\hbox{E}\kern-.125emX}}
\newcommand{\ignore}[1]{}

{%
\setlength{\fboxsep}{0pt}%
\setlength{\fboxrule}{0pt}%
}%

\begin{document}

\title{Stochastic Spiking Attention:
Accelerating Attention with Stochastic Computing in Spiking Networks
}

\author{\IEEEauthorblockN{Zihang Song, Prabodh Katti, Osvaldo Simeone, Bipin Rajendran}
\IEEEauthorblockA{Department of Engineering, King's College London, London WC2R 2LS, U.K.}
\thanks{This work is supported in part by the European Union’s Horizon Europe project CENTRIC (101096379), the EPSRC project (EP/X011852/1) and by  Open Fellowships of the EPSRC (EP/W024101/1 and EP/X011356/1).}
\thanks{Corresponding author: Bipin Rajendran (bipin.rajendran@kcl.ac.uk).}
}


\maketitle

\begin{abstract}
  Spiking Neural Networks (SNNs) have been recently integrated into Transformer architectures due to their potential to reduce computational demands and to improve power efficiency. Yet, the implementation of the attention mechanism using spiking signals on general-purpose computing platforms remains inefficient. In this paper, we propose a novel framework leveraging stochastic computing (SC) to effectively execute the dot-product attention for SNN-based Transformers. We demonstrate that our approach can achieve high classification accuracy ($83.53\%$) on CIFAR-10 within 10 time steps, which is comparable to the performance of a baseline artificial neural network implementation ($83.66\%$). We estimate that the proposed SC approach can lead to over $6.3\times$ reduction in computing energy and $1.7\times$ reduction in memory access costs for a digital CMOS-based ASIC design. We experimentally validate our stochastic attention block design through an FPGA implementation, which is shown to achieve $48\times$ lower latency as compared to a GPU implementation, while consuming $15\times$ less power.   
\end{abstract}


\vspace{-2mm}
\section{Introduction}
The self-attention mechanism at the core of the   Transformer architecture implements a general-purpose form of memory-based processing that can natively encompass multi-modal data and accounts for long-range dependencies \cite{vaswani2017attention}, achieving state-of-the-art (SOTA) performance across a spectrum of tasks \cite{brown2020language,devlin2018bert,sperber2018self,dosovitskiy2021image,ramachandran2019standalone}. However, the standard self-attention block relies on pairwise token operations that have quadratic computational complexity as a function of the number of tokens, thereby significantly increasing computational and memory demands \cite{keles2023computational}. These requirements present notable challenges for the deployment of Transformers in edge AI devices that have stringent constraints on operating power and computational resources  \cite{shazeer2019fast}. While numerous algorithmic approaches have been proposed to achieve sub-quadratic complexity scaling for self-attention \cite{katharopoulos2020transformers,peng2023rwkv}, they invariably introduce computational inaccuracies, potentially leading to reduced accuracy or increased vulnerability to adversarial inputs \cite{keles2023computational}.

\subsection{State of the Art}

There have been recent efforts in designing application-specific integrated circuits (ASICs) for accelerating Transformer architectures that rely on augmenting the level of parallelization, localizing intermediate storage, or quantizing computations \cite{park2020optimus,ham2020accelerating,marchisio2023swifttron}. 
 Another avenue being explored is the application of processing-in-memory techniques, aiming to reduce energy consumption associated with memory access \cite{yang2020retransformer,yang2022fullcircuit}. All these solutions are based on traditional implementations based on real-valued multiply-and-accumulate operations within artificial neural network (ANN) architectures.

Spiking neural networks (SNNs) \cite{maass1997networks} are also being investigated to enhance the efficiency of the self-attention mechanism \cite{she2022spikeformer,zhou2022spikformer,bal2023spikingbert,zhu2023spikegpt}. Distinct from traditional ANNs, SNNs encode data and activations into temporal-coded binary spikes, thereby potentially reducing processing power usage during inference \cite{kulkarni2018spiking}. A primary challenge in this domain is the computation of attention using spike-encoded queries, keys, and values. To address this problem, various SNN-based alternatives to the ANN dot-product attention have been proposed, such as element-wise multiplication and acceptance-weighted key-value methods \cite{yao2023spike, zhu2023spikegpt}. A notable advancement was put forth in \cite{zhou2022spikformer}, which proposed an implementation of the dot-product attention by executing matrix multiplication at each time step, closely approximating the baseline ANN implementation. 

\begin{figure}[t]
\centering
\includegraphics[width=8.1cm]{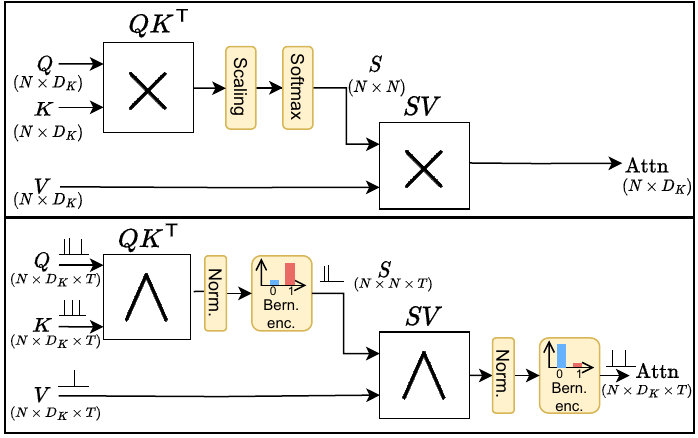}
\caption{Top: A conventional implementation of an attention block based on real-valued multiply-and-accumulate operations within an artificial neural network (ANN) architecture. Bottom: The proposed spiking neural network (SNN)-based attention block with spiking inputs, outputs, and stochastic computations. Multiplication operations are replaced with logical AND $(\land)$ operations on spikes. Further hardware efficiency is achieved by the replacement  of scaling and softmax blocks with a Bernoulli rate  encoder, as discussed in Section \ref{sec:stoattn}.} 
\label{hierarchy2}
\end{figure}

Calculating the dot-product attention with spike-based methods simplifies the complex floating-point (FP) matrix multiplications of the conventional implementation illustrated in Figure \ref{hierarchy2} by turning them into integer multiplications. However, implementing spike-based models on standard CPUs and GPUs generally leads to significant energy inefficiencies. This is due to (\emph{i}) the resource-sharing overhead caused by frequent memory access required for storing and reading the intermediate results of self-attention computations; and (\emph{ii}) the underlying use of high-precision resources in general-purpose computing platforms, encompassing both computation and storage. For instance, CPUs typically operate using 32-bit or 64-bit floating-point (FP32/64) precision for storage and computation. Even though some GPUs are capable of computing as low as 8-bit integers (INT8), this approach is still significantly over-provisioned for handling binary spiking signals. The development of spiking attention accelerators on FPGA devices might be a potential solution, but this area remains under-researched \cite{isik2023hpcneuronet}.

\subsection{Main Contributions}
In this study, we introduce stochastic spiking attention (SSA) -- a novel computing framework implementing a spike-based dot-product attention mechanism. As shown in Figure \ref{hierarchy2}, the SSA design incorporates principles of stochastic computing (SC) \cite{gainestechniques,Alaghi2013Survey}, employing simple AND logic operations for multiplication processes, thereby eliminating the need for complex and power-consuming multipliers, which are commonly utilized in general-purpose platforms. 

The key contributions of the paper are as follows:
\begin{itemize}
\item We introduce SSA, a spike-based attention architecture that builds on SC. SSA is demonstrated to achieve classification accuracy comparable to conventional FP implementations with significantly reduced complexity. Notably, SSA achieves an accuracy of  98.31\% on MNIST and 83.53\% on CIFAR-10 within 10 time steps, which is compared to 99.02\% and 83.66\% respectively for the conventional, ANN-based,  counterpart.
\item We develop a custom hardware accelerator architecture optimized for SSA, employing logical AND gates instead of resource-heavy multiplier units. As compared to a baseline FP ANN implementation, the stochastic SNN attention can lead to a $6.3\times$ reduction in computation energy and a $1.7\times$ reduction in memory access cost.
    
    \item We experimentally validate the SSA block on an FPGA implementation and achieve $48\times$ lower latency and $15\times$ lower power usage than an equivalent GPU implementation.
\end{itemize}


\section{Background}
\subsection{Self-Attention}
For a given input embedding sequence \( X \in \mathbb{R}^{N \times D} \), where $N$ represents the number of tokens and $D$ denotes the embedding dimension, the self-attention block derives the query ($Q$), key ($K$), and value ($V$) matrices through linear transformations as $ Q=XW_Q, K=XW_K, V=XW_V$, where $W_Q$, $W_K$, and $W_V\in \mathbb{R}^{D \times D_K}$ are parameters, with $D_K$ being the key dimension. 
The scaled dot-product attention is calculated as
\begin{equation}\label{eq:selfatt}
    \mathrm{Attn} = \text{softmax}\left( {QK^{\mathsf{T}}}/{\sqrt{D_K}}\right)V,
\end{equation}where the softmax function is applied row-wise. A linear version of self-attention drops the softmax function and was demonstrated to perform comparably with (\ref{eq:selfatt}) at a reduced computational cost \cite{wang2020linformer}.

\subsection{Bernoulli Coding and Stochastic Computing}
Probabilistic, or Bernoulli, coding is an efficient way to translate real values into temporal sequences of independent binary spikes. If $x$ represents the real value to be encoded, the probability $p$ of observing a spike, i.e., a binary 1, at any discrete time step $t$ follows a Bernoulli distribution $p(x^t=1)=\textrm{norm}(x)$, where $x^t$ denotes the probabilistic representation of $x$ at the $t$-th time step and $\textrm{norm}(\cdot)$ denotes a linear normalization function mapping $x$ to the interval $[0, 1]$. We write this encoding operation as \begin{equation}    x^t \sim \mathrm{Bern}(\textrm{norm}(x)),  \end{equation} with variables $x^t$ being independent and identically distributed.

Bernoulli coding supports SC, a computational paradigm that employs Bernoulli bit streams \cite{gainestechniques}. Consider two Bernoulli spike trains $x_{\text{in1}}^t$ and $x_{\text{in2}}^t$ with spiking  rates $p(x_{\text{in1}}^t=1)=\textrm{norm}(x_{\text{in1}})$ and $p(x_{\text{in2}}^t=1)=\textrm{norm}(x_{\text{in2}})$. The multiplication of the original real-valued inputs $x_{\text{in1}}$ and $x_{\text{in2}}$ can be achieved by using a logic AND ($\land$) operation on their stochastic representations, which is denoted as 
\begin{equation}
   x_{\text{out}}^t = x_{\text{in1}}^t \;\land\; x_{\text{in2}}^t,
\end{equation}
since the probability of a spike at time $t$ in the output bit stream is the product  $p(x_{\text{out}}^t=1) = \textrm{norm}(x_{\text{in1}}) \cdot \textrm{norm}(x_{\text{in2}})$. Refer to \cite{Alaghi2013Survey} for a comprehensive survey of SC.


\subsection{Spiking Neurons}
An SNN architecture consists of spiking neurons, whose weights are maintained in real-valued form, while the activations consist of binary signals, i.e., of spikes. Binary coding of activations confers spiking neurons, and thus SNNs, their potential computational efficiency by transforming the standard multiply-and-accumulate (MAC) operations needed for matrix multiplications into simpler accumulation (AC) operations \cite{yao2023spike}. Such accumulation produces the internal state of a spiking neuron, which determines the timing of spike generation via a threshold mechanism. In this paper, we specifically adopt the standard leaky integrate-and-fire (LIF) neuron model \cite{cessac2011discrete}.

\section{Stochastic Spiking Attention}\label{sec:stoattn}
In this section, we introduce the proposed SSA framework, which is illustrated in Figure 2. We detail input coding, self-attention computation, dataflow design, and hardware optimization.

\subsection{Spike Coding of Inputs and Weight Matrices}

To utilize the proposed SSA architecture in Figure 2, we implement a first layer of LIF spiking neurons to evaluate the query, key, and value matrices. This is done as follows. 

First, the input matrix $X$, consisting of the $N$ tokens, is converted element by element into a stochastic bit stream, producing an $N\times D$ binary matrix $X^t$ at each time $t$ over a discrete-time period $1\leq t\leq T$. This conversion uses Bernoulli coding as explained earlier.  For hardware realizations, Bernoulli coding can be implemented using a pseudo-random number generator (PRNG) and comparator.


Then, following the same procedure as in \cite{zhou2022spikformer}, we  generate query \(Q^{t}\), key \(K^{t}\), and value \(V^{t}\) matrices via a layer of LIF neurons. For a $D_K \times N$ matrix sequence $Z^t$ over time $t$, let us write as $\mathrm{LIF}(Z^t)$ the output at time $t$ of a layer of LIF neurons, with one neuron for each entry of matrix $Z^t$. Each LIF neuron takes as input the corresponding real-valued entry of the matrix and produces a binary sequence. 

The spike-encoded input $X^t$ is first multiplied by the matrices $W_Q$, $W_K$, and $W_V$, and then the resulting matrix sequences are fed to LIF neurons, producing the $D_K \times N$ binary outputs
\begin{equation}\label{eq:LIF}
    Q^t = \mathrm{LIF}(X^t W_Q), 
    K^t = \mathrm{LIF}(X^t W_K), 
    V^t = \mathrm{LIF}(X^t W_V).
\end{equation}
While this operation can be accelerated using in-memory computing, in this paper we focus on accelerating the self-attention mechanism block that follows this encoding layer.
\vspace{-3pt}
\subsection{Stochastic Spiking Attention (SSA)}
\vspace{-1mm}

\begin{figure}[t]
    \centering
    \includegraphics[width=8.5cm]{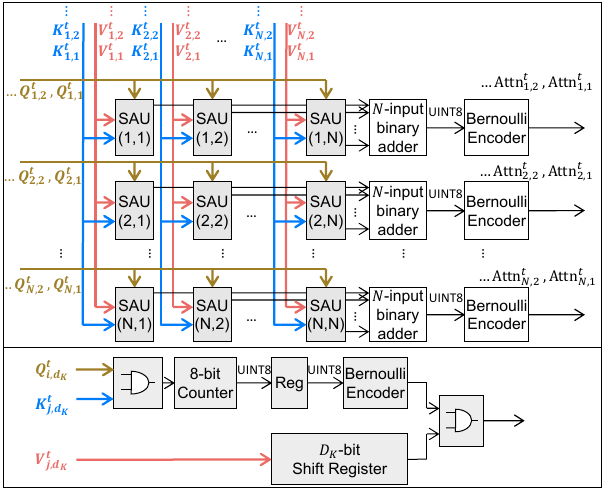}\vspace{-2mm}
    \caption{Top: The architectural schematic of the SSA block. Bottom: the $(i,j)$-th stochastic attention unit (SAU) illustrated in detail. All the wires, unless specified, carry one bit.}\label{fig:stoattn}
    \vspace{-2mm}
\end{figure}

SSA is based on the observation that the binary, stochastic, nature of the encoded sequences \(Q^{t}\), \(K^{t}\), and \(V^{t}\) produced by the LIF layer (\ref{eq:LIF}) supports efficient multiplication using SC, as reviewed in the previous section. SSA leverages this idea to implement a linear form of attention \cite{wang2020linformer} via a sequence of two layers of stochastic computations implementing the query-key product and the attention-value product in (\ref{eq:selfatt}).

Specifically, the $N\times N$ attention scores $QK^\mathrm{T}/\sqrt{D_K}$ in (\ref{eq:selfatt}) are represented via an  $N\times N$  binary matrix sequence \(S^{t}\). To obtain it, the dot-product operation between the entries of the binary matrix sequences \(Q^t\) and \(K^t\) is first evaluated via SC by summing the results of logic AND operations across the dimension $D_K$. For each $(i,j)$-th entry, this yields the real number  $\sum_{d_k=1}^{D_K} Q^t_{i, d_k} \land K^t_{j, d_k}$, which upon  normalization  is used for Bernoulli encoding as \begin{equation}\label{eq:attentionscore}
    S^t_{i, j} \sim \mathrm{Bern}\left(\frac{1}{D_K}\sum_{d_k=1}^{D_K} Q^t_{i, d_k} \land K^t_{j, d_k}\right).
\end{equation}

Finally, the attention-value product $(QK^{\mathsf{T}}/\sqrt{D_K})$  is evaluated in a similar way via a cascade of SC and Bernoulli encoding. This yields the $N\times D_K$ binary matrix sequence 
\begin{equation}\label{eq:weightsum}
    \mathrm{Attn}^t_{i,d_k} \sim \mathrm{Bern}\left(\frac{1}{N}\sum_{j=1}^{N} S^t_{i, j}\land V^t_{j,d_k}\right)
\end{equation} for discrete time $t=1,...,T$.

The proposed SSA architecture can be incorporated into an overall Transformer architecture for end-to-end training using standard surrogate gradient methods for SNNs \cite{neftci2019surrogate}. 
\begin{figure*}
    \centering
    \includegraphics[width=0.9\textwidth]{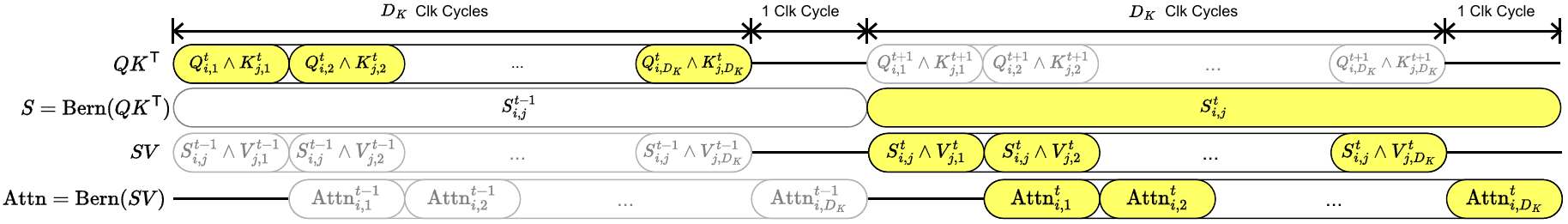}\vspace{-2mm}
    \caption{Illustration of dataflow design for the attention operation of each  $(i,j)$-th stochastic attention unit (SAU).}
    \vspace{-2mm}
    \label{fig:dataflow}
    \vspace{-5mm}
\end{figure*}

\subsection{Dataflow Design for SSA}
To implement the SSA computations specified in \eqref{eq:attentionscore} and \eqref{eq:weightsum}, the proposed architecture leverages stochastic attention units (SAUs)  arranged in an SSA block as depicted in Figure \ref{fig:stoattn}. Since the computation of the attention score $S^t$ at each time step $t$  requires $N^2D_K$ operations, our design employs  $N^2$ physical SAU units arranged as an \(N \times N\) array, and we compute $S^t$ in $D_K$ clock cycles. The dataflow is illustrated in Figure \ref{fig:dataflow}. 

The \(N^2\) parallelization level in SAUs offers significant benefits. First, it enables simultaneously sending each query to each key-value pair. This is achieved by simultaneously streaming \(K^t\) and \(V^t\) across rows and  \(Q^t\) across columns. Furthermore, it allows for the simultaneous generation of all $N\times N$ elements of the attention score \(S^t\) matrix. The \((i,j)\)-th element of matrix \(S^t\) is computed within the corresponding \((i,j)\)-th SAU.  Thus, the $N^2$ parallelization of SAUs enables computations in  \eqref{eq:attentionscore} for each $(i,j)$-th pair to be conducted in parallel. This efficient data streaming through the SSA block eliminates the need for writing/reading intermediate data from the memory. Since SOTA Transformers for edge AI applications have $N$ in the range of $16-128$, we have employed $N^2$ physical SAU units in our design.


To evaluate the attention weights via  \eqref{eq:attentionscore}, the \(D_K\) AND operations are executed serially in time. The summation is realized by counting the AND output using a counter with UINT8 output, accommodating a key dimension \(D_K\) up to \(2^8 = 256\). After every \(D_K\) clock cycles, the summation is buffered to a Bernoulli encoder, where it is normalized by $D_K$ and used as a probability to generate the Bernoulli sample $S^t_{i,j}$. 


To calculate the self-attention output  \eqref{eq:weightsum}, the generated attention weights $S^t_{i,j}$ are held for $D_K$ clock cycles in each SAU. The AND result between $S^t_{i,j}$ and $V^t_{j,d_K}$ is calculated using another AND gate, serving as the output of the SAU. The summation in \eqref{eq:weightsum} is achieved by adding the outputs of the SAUs in each row using an $N$-input binary adder. The sum is then sent to another Bernoulli encoder, where it is normalized by $N$ and used as a probability to generate the Bernoulli sample $\mathrm{Attn}^t_{i,d_k}$. $\mathrm{Attn}^t_{i,d_k}$ are generated sequentially by holding $S^t_{i,j}$ in the $(i.j)$-th SAU when streaming $V^t_{j,d_K}$, where $d_K$ runs from 1 to $D_K$.



As a result of this procedure, the \(i\)th row of SAUs within the SSA block sequentially outputs the values \(\mathrm{Attn}^t_{i,1}\), \(\mathrm{Attn}^t_{i,2}\), ..., \(\mathrm{Attn}^t_{i,D_K}\). With \(N\) rows of SAUs operating in parallel, the entire matrix \(\mathrm{Attn}^t\) is acquired column by column. 

\subsection{Optimizing for Hardware Implementation}
We conclude this section with notes on how to optimize the proposed SSA architecture for hardware implementation. First, the step \eqref{eq:weightsum} requires the value $V^t$ to be streamed to the AND gate when $S^t$ becomes valid. To avoid using external delays for $V^t$, a \(D_K\)-bit shift register operating on a first-in-first-out basis is deployed in each SAU to temporarily buffer \(V^t\) and align it with $S^t$. This allows for the simultaneous streaming of $Q^t$, $K^t$, and $V^t$ within the SSA block, facilitating pipelining over time steps. Second, our design employs linear feedback shift registers-based PRNGs to generate the random number within Bernoulli encoders. To improve power and area efficiency, we have employed a custom reuse strategy for random number generation, similar to \cite{joshi2020essop}.  
Third, selecting $D_K$ and $N$ as powers of two can further streamline the hardware design by eliminating the need for normalization before the operation of Bernoulli encoders. This simplification allows the Bernoulli samples to be calculated through a direct comparison between the sum and a fixed-point random integer sampled from a uniform distribution.
\vspace{-5pt}

\section{Experiments}\label{se_exp}
We evaluated three implementations of the ViT-Small model (composed of 6 encoder layers and 8 attention heads per layer) on the MNIST and CIFAR-10 datasets to assess their performance in image classification tasks: 
(\emph{i})  {ANN} -- a SOTA ANN accelerator as reported in \cite{marchisio2023swifttron};
(\emph{ii})  {Spikformer SNN} -- an SNN with its dot-product attention implemented with integer multipliers \cite{zhou2022spikformer}; 
and (\emph{iii}) the proposed SNN-based  {SSA} architecture.
The parameters of all three implementations are INT8-quantized. The ANN was tested with INT8-quantized activations, while the SNNs were evaluated with binary activations at 4, 8, and 10 time steps. As shown in Table \ref{tb:acc}, SSA ViT achieves a peak accuracy of $98.31\%$ on MNIST and $83.53\%$ on CIFAR-10, within 10 time steps, which is comparable to that of the baseline ANN ViT implementation.  


\begin{table}[t]
\centering
\caption{Comparison of classification accuracies for different ViT-Small architectures on MNIST and CIFAR-10 datasets.}\label{tb:acc}
\begin{tabular}{|K{1.7cm}|c|K{2.0cm}|K{2.0cm}|}
\hline
 Architecture      & $T$ & Accuracy -- MNIST       & Accuracy -- CIFAR-10    \\ \hline 
   ANN   &    -   & 99.02             & 83.66             \\ \hline
Spikformer   & 4/8/10     & 98.17/98.21/98.34 & 83.32/83.4/83.41  \\ \hline
SSA & 4/8/10     & 97.83/98.17/\textbf{98.31} & 81.57/83.31/\textbf{83.53} \\
\hline\end{tabular}
\end{table}

We estimate the energy consumption for the attention block for the three architectures by accounting for all the required compute and memory access (read/write) operations, following the approach  in \cite{acesnn2022}. The calculations assume the basic energy metrics for 45 nm CMOS technology as reported in \cite{pedram2017dark,buffa2021voltage}. We assume that all required data for computations are available on the on-chip static random-access memory (SRAM). 

As shown in Table \ref{tb:energy}, for a single attention block, SSA exhibits $6.3\times$ and $5\times$ reduction in processing energy consumption compared to ANN attention and Spikformer attention (measured over 10 time steps), respectively.  Since memory access occurs at each time step, SNNs do not exhibit any advantage over 8-bit quantized ANNs for memory accesses. We observed that in 10 time steps, Spikformer attention consumes more energy for memory access than INT8-ANN attention. The SSA architecture, in contrast, reduces memory access energy by $1.7\times$ compared to ANN and $1.9\times$ relative to Spikformer attention. Overall, our implementation demonstrates an improvement in total energy efficiency by $1.8\times$ and $2\times$ when compared to ANN   and Spikformer attention.

\begin{table}[t]
\centering
\caption{Comparison of total (processing + memory) energy consumption for a single attention block of three different architectures. For SNN models,  $T=10$ time steps.}\label{tb:energy}
\begin{tabular}{|K{2.5cm}|K{1.38cm}|K{1.88cm}|K{1.35cm}|}
\hline
Architecture                & Processing Energy ($\mu$J)   & Memory Access Energy  ($\mu$J)     &  Total Energy  ($\mu$J)   \\ \hline 
ANN Attention              & 7.77                         & 89.96                              &  97.73                  \\ \hline
Spikformer Attention     & 6.20                         & 102.85                             &  109.05                     \\ \hline
SSA                     & \textbf{1.23}                & \textbf{52.80}                     &  \textbf{54.03}             \\ \hline
\end{tabular}
\end{table}
\vspace{-3pt}
\begin{table}[t]
\centering
\vspace{-2mm}
\caption{Comparison of hardware efficiency for a single attention block (CPU and GPU implementations) for different architectures. For SSA block, $T=10$. }\label{tab:fpgavsgpu}\vspace{-2mm}
\begin{tabular}{|K{2.8cm}|K{1.3cm}|K{1.6cm}|K{1.3cm}|} 
\hline
Architecture -- Device    & $f_{\text{clk}}$ (MHz)    & Latency (ms)  & Power (W)     \\ \hline 
ANN attention -- CPU     & $2100$                      & $0.15$                    & $107.01$                  \\ \hline
ANN attention -- GPU     & $562$                       &$0.06$                      & $26.13$                  \\ \hline
SSA -- CPU           & $2100$                      & $2.672$                     & $65.54$                   \\ \hline
SSA -- GPU           & $562$                       & $0.159$                     & $22.41$                 \\ \hline
SSA -- FPGA          & $200$                       & $\mathbf{3.3\times 10^{-3}}$            & $\mathbf{1.47}$         \\ \hline
\end{tabular}
\end{table}



We also implemented the proposed SSA block on a lightweight FPGA (within Xilinx Zynq-7000 SoC). 
The latency and power consumption of the FPGA implementation were compared against CPU (Intel i7-12850HX) and GPU (Nvidia RTX A2000), as shown in Table \ref{tab:fpgavsgpu}. We obtain $48\times$ better latency expending $15\times$ lesser power than that of the GPU implementation. We also note that the SSA block on FPGA has $18\times$ less latency while consuming $17\times$ less power than the ANN GPU implementation.

\vspace{-2mm}
\section{Conclusion}
This study presents a stochastic spiking attention (SSA) framework and its hardware design. We demonstrate that the proposed  SSA framework achieves close to baseline ANN accuracy for image classification tasks with ViT-Small model, with over $1.8\times$  estimated gains in energy efficiency. 
Furthermore, an FPGA implementation of an SSA block shows $48\times$ lower latency while consuming $15\times$ less power than GPUs. Overall, this work supports the implementation of Transformer-based models for energy and resource-constrained applications,  including mobile devices in 6G.

\bibliography{reference}

\end{document}